\documentclass{aa}  
\usepackage{graphicx}
\usepackage{txfonts}
\usepackage[utf8]{inputenc}
\usepackage{orcidlink} 

\newcommand{\be}{\begin{equation}}
\newcommand{\ee}{\end{equation}}
\newcommand{\rund}[1]{\left(#1\right)}
\newcommand{\vc}[1]{\mbox{\boldmath $#1$}}

\newcommand{\eck}[1]{\left[ #1 \right]}

\begin{document} 
\title{The polarization of strongly lensed point-like radio sources}
\author{Xinzhong Er
      \inst{1}\orcidlink{https://orcid.org/0000-0002-8700-3671}  }
\institute{Tianjin Astrophysics Center, Tianjin Normal University, Tianjin 300387, P.R.China
\email{phioen@163.com}
 }
\date{Received; accepted }
\abstract
{} 
{The magnetized medium induces birefringence, splitting the light into two distinct wave modes. The differing propagation speeds of the two modes result in different trajectories. Strong gravitational lensing amplifies the birefringence and introduces an additional geometric rotation on top of the Faraday rotation. We compare the geometric rotation with the Faraday rotation.}
{We construct the lens equation for massive objects in a magnetized plasma environment, and calculate the time delay difference between the two modes using two toy examples. We present that in the strong lensed radio sources, birefringence causes geometric rotation, which is a non-negligible effect, even with a weak magnetic field. }
{In both examples, the geometric delay causes a comparable or stronger rotation than the Faraday rotation and show a similar dependence on the wavelength of the signal. For a point lens with a strong magnetic field, the two wave modes exhibit distinct behaviours. The polarization of lensed sources can provide additional insights into the magnetic field and plasma environment.}
{}
\maketitle
%
\section{Introduction}
Strong gravitational lensing serves as a powerful tool in astrophysics, enabling the estimation of mass distribution without assuming the nature or status of the matter. This allows us to constrain the model of formation and evolution of the galaxy and galaxy cluster \citep{1996astro.ph..6001N,2006glsw.conf....1S}. In lensing, multiple images can form when a background source is well aligned with the lens. The propagation of light along different trajectories introduces a relative time delay between the images, resulting from both the Shapiro effect and the difference in path length. It has been found that the lensing time delay can be used to estimate the Hubble constant \citep{1964MNRAS.128..307R}, providing a measurement independent of the cosmic distance ladder or the Cosmic Microwave Background. Constraints have been carried out from observations \citep[e.g.][]{2013A&A...556A..22T,2014ApJ...788L..35S}. In order to measure the time delay between multiple images, the background source must exhibit variability. Quasi-stellar objects (QSOs) are commonly used for this purpose due to their high number density in the universe. However, because of their stochastic variability and long time scale, it requires extensive monitoring to achieve an accurate measurement of the time delay. 
Fast radio burst (FRB) is a class of extragalactic radio transient with millisecond duration \citep[e.g.][]{2007Sci...318..777L,frbreview2019}. Thanks to its short duration and high brightness, FRB offer a promising alternative for the background source in time delay measurement, and to study cosmology and fundamental physics \citep[e.g.][]{2021A&A...645A..44W}. 

In addition to gravity, plasma can deflect light as well \citep[e.g.][]{2015PhRvD..92j4031P}. The influence of plasma lensing depends on the frequency of light. Given the typical density of plasma in the interstellar or intergalactic medium ($\sim10^{-5}-10^{-1}$ electron cm$^{-3}$), the effects of plasma lensing are minimal and detectable primarily at low frequencies \citep[e.g.][]{2010MNRAS.404.1790B,2016ApJ...817..176T}. The deflection angle caused by plasma is on the order of 1 milli-arcsec and generally negligible concerning the image positions. However, variations in magnification/brightness and time delay can be significant and detectable, especially for the short-duration signal. For example, strong brightness variations observed in one repeating FRB have been ascribed to a plasma lensing event \citep{2024MNRAS.531.4155C}. 
Moreover, polarization measurements provide critical insights into the emission mechanism of the sources and the properties of magnetic field along the line of sight. When light propagates through a magnetized plasma, the two circular polarization modes experience different refractive indices due to their distinct propagation speeds. This phenomenon leads to rotation of the linear polarization of the signal, known as the Faraday rotation. Plasma lensing in the presence of strong variations in the magnetic field can exhibit birefringence effects, which may produce intriguing observational signatures. For example, when the lensed image lies close to the critical curves, the polarization state of the background source can even undergo a flip \citep{2019ApJ...870...29S,2019MNRAS.486.2809G,2023MNRAS.522.3965E}. In extreme cases, the separation of the two polarization modes can result in splitting of the lensed images.

Birefringence can also occur in strongly lensed radio sources, with even cross-talk between  gravitational lensing and plasma lensing. The propagation of light through magnetic plasma in a gravitational field has been explored in several studies \citep[e.g.][]{1980RSPSA.370..389B,1981A&A....96..293B,1981RSPSA.374...65B,2003MNRAS.342.1280B}. The dispersion relation exhibits complex behaviour in the presence of a gravitational field. In case of a strong magnetic field, the lensed images of the two modes can even split as well \citep{2019IJMPD..2840013T}. In this work we investigate the birefringence in strongly lensed radio sources, focusing on magnification and time delays. Additional lensed images can form, and importantly, the distinct propagation paths of the two polarization modes lead to different arrival time. As a result, the Faraday rotation relation cannot be used as an accurate estimate of the magnetic field.
We demonstrate these effects using two toy examples: a galaxy scale lens with a weak magnetic field, the other one is a point lens with a strong magnetic field.

\section{The deflection with magnetic field}
\label{sec:formulae}
The influence of the magnetic field on the propagation of light in curved space-time will be studied. We explore several astrophysical scenarios, such as an FRB lensed by a galaxy or a galaxy cluster, where the magnetic field is relatively weak. Additionally, we consider cases where light propagates near compact objects, such as neutron stars or black holes, which are associated with strong magnetic field. In both scenarios, we restrict our analysis to static fields, i.e., there is no temporal variation in the gravitational field, plasma density, or magnetic field. 

In general, solving the Maxwell equations in a relativistic medium exactly is not feasible. Therefore, several approximations are employed in this work. First, we adopt the framework of Hamilton's geometric theory of rays to study the optics in a curved space-time with a magnetized medium. Secondly, we restrict our study to the weak gravitational field and small deflection angles, i.e. focusing exclusively on the far-field regime. For scenarios involving light deflection near black hole or neutron star, e.g. to study the shadow of the black hole, one can find more details in e.g. \citet{2022PhR...947....1P}. We apply the thin lens approximation, as the lens size is small compared to the distance between the lens and the observer. 
Moreover, we model the electromagnetic waves as as plane, and monochromatic, with wavelength significantly smaller than the characteristic scale of plasma variation \citep{1980RSPSA.370..389B,2003MNRAS.342.1280B}. Thus, the wave optical effects are not considered in this analysis \citep[e.g.][]{2020MNRAS.497.4956J,2023MNRAS.520.2995F,2024MNRAS.534.1143S}. 

We denote plasma frequency by $\omega_e$\footnote{In this work, we only consider free electrons in the plasma. The ions can slightly change our result, but for low temperature, the corrections are small \citep[$\sim0.03\%$][]{1975JPlPh..13..571I}.} and Larmor frequency by $\omega_B$
\be
\omega_e^2(x) = \frac{4\pi \, n_e(x)\, e^2}{m_e}
\equiv K_e n_e(x), \quad \omega_B(x)=\frac{e}{m_e}\vc{B}(x)\cdot\vc{\hat{k}},
\ee
where $n_e(x)$ represents the number density of electron in three dimensions, $\vc{B}(x)$ is the magnetic field, and $\vc{\hat{k}}$ is the unit wave vector. Typically, both the plasma frequency and the Larmor frequency are low. For example, a plasma with electron density $n_e=1000$\,cm$^{-3}$ has a frequency of only $\sim0.3$ MHz, and a magnetic field of $B=1$ gauss yields a Larmor frequency about $2.8$ MHz. These values are significantly lower than the typical observational frequencies ($\sim$ GHz), ensuring that $\omega\gg \omega_e,\omega_B$ for most observations.

\subsection{Nearly flat spacetime}
The dispersion relation is critical for this study particular since the effect of the magnetic field in curved spacetime is complex. We adopt the results from \citet{2003MNRAS.342.1280B,2004MNRAS.349..994B}, which use the locally flat centre-of-mass rest frame. The function $D(k_\mu,x^\mu)$ is analogous to the Hamiltonian of the wave in the magnetic plasma \citep{1980RSPSA.370..389B,1981RSPSA.374...65B}. We consider the far field, where the space can be approximated as nearly flat. We also apply the quasi-longitudinal approximation for the cold plasma, resulting in transverse wave modes. As a consequence, certain modes are absent when the light is deflected. However, this inaccuracy is acceptable for cases only involving small deflection angles (more discussion later). The dispersion relation is similar to that in the flat spacetime
\be
D=\eck{g^{\mu\nu} k_\mu k_\nu +
\omega_e^2 \pm \omega_e^2\frac{\omega_B}{\omega(x)} },  \label{eq:dispersion-flat}
\ee
We emphasize that the observational frequency $\omega(x)$ is not a constant, but depends on the gravitational field, i.e. $g_{00}$\footnote{gravitational redshift}. We neglect such variations since we focus on the weak field cases.
Then we can solve for the ray equations by
\be
\frac{dx^\nu}{d\tau}=\frac{\partial D}{\partial k_\nu},\quad
\frac{d k_\nu}{d \tau} = -\frac{\partial D}{\partial x^\nu},
\label{eq:dis-ray1}
\ee
where $\tau$ is an affine parameter. The refractive indices can be written as 
\be
n_{L,R} = 1- \frac{\omega \omega_e^2}{2\omega^2(\omega \pm \omega_B)}
\approx 1- \frac{\omega_e^2}{ 2\omega^2}\rund{1\pm \frac{\omega_B}{\omega}},
\label{eq:refractive-index1}
\ee
where subscript $L,R$ stands for left and right mode of circular polarization. Throughout the rest of the paper, We will omit the subscript $L,R$. The approximation in Eq.\,\ref{eq:refractive-index1} is valid under the condition $\omega_B\ll \omega$.
The trajectories of the rays can be determined by integrating Eqs.\ref{eq:dis-ray1}. We consider a stationary, spherically symmetric plasma distribution with a Schwarzschild metric. Without loss of generality, we focus on the light ray confined to the plane of $\theta=\pi/2$, the metric can be written as 
\be
ds^2 = -(1-\frac{2GM}{r})\, dt^2 + (1-\frac{2GM}{r})^{-1}dr^2 + r^2 d\phi^2 .
\ee
Since $g_{\alpha\beta}=0$ for $\alpha\neq\beta$, then we can obtain the equation
\begin{align}
\frac{dk_\nu}{d\tau}& = \frac{1}{2} \eck{g^{\nu\nu}_{,\nu}k_\nu^2+(\omega_{e}^2)_{,\nu} \pm (\omega_e^2\omega_B/\omega)_{,\nu} }
\label{eq:vector-deri}
\end{align}
The last term in the equation represents the influence of the magnetic field, introducing anisotropy into the system. For wave propagating perpendicular to the magnetic field, i.e. $\vc{B}\cdot \vc{k}=0$, the magnetic field has no effect on the propagation of the wave. However, for waves aligned parallel or antiparallel to the magnetic field, the field will exert opposing effects on the propagation, a consequence of the distinct behaviours of the left- and right-hand mode polarizations. As light propagates, its direction changes due to the deflection, causing the ordinary and extraordinary waves to interchange along the path. This results in a varying contribution from $\vc{B}\cdot \vc{k}$. In this study, we restrict our analysis to the small deflection angle and neglect this effect. We acknowledge that this approximation breaks down at a large deflection angle, such as those encountered near black holes or neutron stars. Thus, the scope of this work is only limited to weak field scenarios, and we assume that the magnetic field remains either parallel or antiparallel to the wave vector throughout the propagation. 
For high frequencies, two mode waves propagate approximately along the same rays but with distinct speeds. This approximation is crucial for the phenomenon of Faraday rotation. In this work, we investigate the subtle differences in the ray paths of the two modes and the corresponding effects.

We use $e^i$ and $e_i$ for the unit vectors. The components of the vector $k^\nu$, in a non-homogeneous medium can be written as
\begin{align}
k^{\nu} &= (\omega, n\omega e^i), \nonumber \\
k_{\nu} &= (-\omega, n\omega e_i).
\end{align}
With the small angle approximation, we can select the z-axis for the trajectory of the light.
From Eq.\,\ref{eq:vector-deri},
\be
\frac{d e_{\nu}}{dz} = \frac{1}{2} \rund{g_{33,\nu} +\frac{1}{n^2}g_{00,\nu} -\frac{(\omega_{e}^2)_{,\nu} \pm (\omega_e^2\omega_B/\omega)_{,\nu}}{n^2\omega^2} },
\label{eq:alpha-define}
\ee
where $\nu=1,2$. We further assume that the plasma distribution around the lens is spherically symmetric, i.e. $n_e(r)$. This assumption can be relaxed to axial symmetry if necessary. By introducing the impact parameter $b$, defined as the separation between the lens and the photon trajectory, the position of the photon can be characterized by $b$ and $z$ along the line of sight, such that $r=\sqrt{b^2+z^2}$. The deflection angle is then derived by an integral along the direction of the ray $e_\alpha$ 
\citep{2003MNRAS.342.1280B,2010MNRAS.404.1790B}, and $\hat{\alpha} = e_\alpha(+\infty)-e_\alpha(-\infty)$, 
\be
\hat{\alpha} =\frac{1}{2}\int_{-\infty}^{\infty}
dz\, \frac{b}{r}\eck{g_{33,r} + \frac{1}{n^2}g_{00,r} - \frac{K_e n_{e,r}}{n^2\omega^2} \pm \frac{K_eK_b (n_eB)_{,r}}{n^2 \omega^3}}. 
\label{eq:def-ang-hxy}
\ee
Here we use $\omega_B = e B(r)/(m_e)\equiv K_b B(r)$.  

The first term represents the gravitational deflection. The second accounts for the deflection caused by a homogeneous plasma, which is typically small and will be neglected in this work. After reorganizing the equation, we express the total deflection angle as a combination of three parts
\begin{align}
\hat{\alpha}&= \hat{\alpha}_{gl} + \hat{\alpha}_{ pl} \pm \hat{\alpha}_B,\\
\approx &-\frac{2R_s}{b} -  \frac{K_e}{2} \int \frac{n_{e,\alpha}}{\omega^2-\omega_e^2} dz \pm \frac{K_e K_b}{2} \int 
\frac{(n_eB)_{,\alpha}}{\omega^3} \,dz.
\label{eq:def-all}
\end{align}
where $\hat{\alpha}_{gl}$ and $\hat{\alpha}_{pl}$ is the deflection angle caused by gravity, and the density gradient of the plasma, i.e the refraction respectively. The third term $\hat{\alpha}_B$ is a deflection generated by the gradient of the magnetic field. Usually, $\alpha_{B} < \alpha_{pl}\ll \alpha_{gl}$ in lensing systems, and $\alpha_B$ is negligible. However, we retain this term in our analysis since it introduces a critical distinction between the two polarization modes. 
Higher-order contributions arise from Eq.\,\ref{eq:alpha-define}, which caused by the variations in the observational frequency along the line of sight as we discussed, are omitted for simplicity. Furthermore, we expand the deflection angle in the final term of  Eq.\,\ref{eq:def-all} to facilitate our analysis
\be
\hat{\alpha_B} = \frac{K_e K_b}{2}\int \left( \frac{n_{e,\alpha} B + n_e B_{,\alpha}}{\omega^3}\right) dz.
\ee
The first contribution arises from variations in the electron density, while the second one from variations of the magnetic field. Thus, even if either the electron density or magnetic field remains constant, a scenario that is unlikely, there still will be a difference between the two polarization modes.

\subsection{Warm plasma}
For the warm plasma, the dispersion relation will be changed by the temperature of the plasma \citep{2003MNRAS.342.1280B},
\be
D= \rund{1+\frac{1}{3}\frac{kT \omega_e^2}{m_e\omega^2}}k^{\mu}k_{\mu} + \omega_e^2 \pm\omega_e^2 \frac{\omega_B}{\omega},
\ee
where $T$ is the temperature of a thermal electron distribution. The alternation in the deflection properties is mainly governed by the temperature of the ionized gas. The gas with high temperature, e.g. $\sim$kev, can emit substantial high-energy radiation, which is directly observable. Thus, if we consider the case that the observational frequency exceeds the plasma frequency, the thermal modification for the cold or warm plasma is found to be less than $0.1\%$. Under such conditions, this effect can be safely neglected. 

\subsection{General cold plasma}
For the general case, the dispersion relation becomes more complex and incorporates the covariant form of the magnetic field \citep{2003MNRAS.342.1280B}. Here we only discuss the dispersion relation in general but still limited to the approximations that we mentioned in previous
\begin{align}
D=&k_\mu k^\mu-\delta\omega^2 -\frac{\delta}{2(1+\delta)}\left\{ \eck{ \omega_B'^2 - (1+2\delta) \omega_B^2 }   \right.
\nonumber\\
&\pm \left. \sqrt{ \omega_B'^4+2( 2\omega^2-\omega_B^2 - \omega_e^2) \,\omega_B'^2 +\omega_B^4} \right\} ,
\label{eq:bb2003dispersion}
\end{align}
where 
\be
\delta=\frac{\omega_e^2}{\omega_B^2-\omega^2}, \quad \omega_B'= \frac{e{\cal B}^\mu k_\mu}{m_e \omega}.
\ee
Here $\omega_B'$ differs from $\omega_B$, especially when light propagates in a strong gravitational field.
We rewrite Eq.\,\ref{eq:bb2003dispersion} in a compact form
\be
D=k_\mu k^\mu-\delta\omega^2 -\frac{\delta}{2(1+\delta)} \eck{A\pm \sqrt{C}},
\label{eq:dispersion-compact}
\ee
where 
\begin{align}
A=&\omega_B'^2 - (1+2\delta) \omega_B^2,\\
C=&(\omega_B'^2-\omega_B^2)^2 +2(2\omega^2 -\omega_e^2)\omega_B'^2.
\end{align}
The parameter $C$ characterizes the splitting of the two polarization modes. The prefactor $\frac{\delta}{2(1+\delta)}$ reveals an intriguing behaviour: when the magnetic field becomes sufficiently strong ($\omega_B^2>\omega^2-\omega_e^2$), the two modes flip. For instance, in the case of a double Einstein ring formed by gravitational lensing with a strong magnetic field, an increase in magnetic field strength can cause the ordinary mode to transition from the inner ring to the outer ring, or vice versa. However, as we will demonstrate later, the conditions required for such image splitting demand an extremely strong magnetic field, making this phenomenon unlikely to be observed in reality.

We further evaluate and simplify the parameters $A$ and $C$ within a Schwarzschild metric, starting from the frequency by
\be
\omega_B'\approx \frac{e B}{m_e (1-2M/r)},
\ee
wherein only the radial component of the magnetic field is considered. In case of a strong gravitational field, e.g. near a pulsar or a black hole, i.e. $r\sim10r_g$, the discrepancy between $\omega_B$ and $\omega_B'$ can reach $\sim10\%$. In contrast, in typical gravitational lensing scenarios, where the deflection angle measures on the order of $\sim1-10$ arcsec, and $r>10^{4}r_g$, the discrepancy becomes negligible.  
In this study, we focus on weak gravitational fields, thus approximate $\omega_B'$ by $\omega_B$. The dispersion relation can then be written as
\be
D= k_\mu k^\mu-\delta\omega^2 +\frac{\delta}{2(1+\delta)}\eck{2\delta \omega_B^2 \mp \sqrt{2(2\omega^2 -\omega_e^2) \omega_B^2} }.
\label{eq:dispersion-app}
\ee
This relation reduces to Eq.\ref{eq:dispersion-flat} in the limit of a weak magnetic field.

\section{Toy models in weak magnetic field}
To illustrate the effect of lensing on polarization, we begin by examining a simple case where the wave vector is parallel to the magnetic field. To understand how lensing modifies polarization, we construct the basic equation of lensing, incorporating the influence of the magnetic field. Our approach follows the general framework of lensing, which includes contributions from both gravitational and plasma effects \citep[e.g.][]{2006glsw.conf....1S,2016ApJ...817..176T,2022MNRAS.516.2218E}. Analogous to lensing in vacuum, we construct a lens potential and calculate the time delay, which consists of both geometric and potential/dispersion contributions. The lens equation connects the angular positions between the image position $\theta$ and source position $\beta$ by the reduced deflection angle $\alpha$
\be
\beta =\theta-\alpha(\theta)
=\theta - \nabla_\theta \psi_g(\theta) 
- \nabla_\theta \psi_{pl}(\theta) -\nabla_\theta \psi_B(\theta),
\ee
where $\nabla_\theta$ is the gradient on the image plane, and $\psi_{g}$, $\psi_{pl}$ is the effective lens potential for gravitational or plasma lensing respectively. All the lensing distortions can be calculated from potential $\psi$. For example, the magnification produced by a lens is inversely proportional to the determinant of the Jacobian matrix $A$ of the lens equation, $\mu^{-1}=$ det$(A)$. The reduced deflection is given by $\alpha=\hat{\alpha}D_{ds}/D_s$, where we use the angular diameter distances between the lens and us, the source and us, and between the lens and the source as $D_d,D_s,D_{ds}$ respectively.
We further define a time delay distance
\be
D_t=(1+z_d) \frac{D_d D_s}{D_{ds}},
\ee
where $z_d$ is the redshift of the lens.
Then the lens potential of the plasma and magnetic deflection can be written as
\be
\psi_{pl} + \psi_B= \frac{c}{D_t}\rund{ \frac{K_e}{2\omega^2}\int n_e dz \pm  \frac{K_eK_b}{2\omega^3} \int n_e B dz}.
\ee
In analogy to the Shapiro delay in the gravitational lensing, there is a delay of the signal with respect to that in the vacuum, which is dubbed as ``potential delay'' or ``dispersion delay'',
\be
t_{pot} = \frac{D_t}{c} (\psi_{pl} +\psi_B).
\label{eq:pot2dt}
\ee
The delay arises essentially from the change of propagation speed within the plasma medium. As this speed depends on frequency, the phenomenon is commonly described by the dispersion relation.
The total time delay including the deflection (the geometric delay) is
\be
t=\frac{D_t}{2c} \alpha^2 +t_{gl} + t_{pl} \pm t_B.
\ee
Usually the geometric delay is negligible compared to the potential delay. However, in case of strong lensing, the geometric delay causes significant contribution \citep{2020CQGra..37t5017T}.
Next, we calculate the delay difference between left and right modes of polarization, as this is our primary interest
\begin{align}
\Delta t_{LR}=&\frac{D_t}{2c}\rund{\alpha_L^2-\alpha_R^2}+2t_B \nonumber \\
=&\frac{2D_t}{c}\alpha_{gl}\alpha_B + \frac{2D_t}{c}\alpha_{pl}\alpha_B + \frac{K_eK_b}{\omega^3}\int n_e B dz.
\label{eq:3timedelay}
\end{align}
The last term arises from the constant magnetic field and leads to Faraday rotation. We further assume that the left- and right-mode polarizations propagate along the same trajectory, implying that the two rays are coherent. Then the phase difference can be convert to the rotation of the linear polarization
\begin{align}
\Delta\phi &= 2\pi \omega \Delta t_{LR} \nonumber \\
&= \frac{4\pi \omega D_t}{c}\alpha_{gl}\alpha_B + \frac{4\pi \omega D_t}{c}\alpha_{pl}\alpha_B + \frac{2\pi K_e K_b}{\omega^2} \int n_e B dz.
\label{eq:3rotations}
\end{align}
Since $\alpha_{gl}$ is achromatic, the first term scales with $\lambda^2$, analogous to the wavelength dependence in Faraday rotation. The second term has a stronger dependence, scaling with $\lambda^4$. While the deflection angles in both geometric terms are generally small, two geometric delays, which are proportional to the distance $D_t$, become significant when the lens is located at cosmological distances. We compare the three terms in Table \ref{tab:phi_depend}. 
\begin{table}[]
\centering
\caption{The redshift and wavelength dependence of $\Delta t$ (Eq.\,\ref{eq:3timedelay}) or the phase rotation for the linear polarization (Eq.\,\ref{eq:3rotations}).}
\begin{tabular}{c||c|c|c|}
   &$\alpha_{gl}\alpha_B$  &$\alpha_{pl}\alpha_B$  &Faraday \\
\hline
wavelength $\lambda$     &2  &4  &2  \\
redshift $z$     &1   &1    &0  \\
\end{tabular}
\label{tab:phi_depend}
\end{table}

We adopt a Singular Isothermal Sphere \citep[SIS][]{2008gady.book.....B} for the mass model of the lens and demonstrate the magnetic effects in a galaxy-scale lens. A moderate redshift for the lens ($z_d=0.2$), and a possible redshift for the background source ($z_s=0.5$) is used, which is suggested by the peak redshift of the FRB \citep{2021ApJS..257...59C}. We use a massive lens galaxy with $\sigma_v=300km/s$, which gives us an Einstein radius of $\theta_E\sim1.1$ arcsec. Similarly, for the density profile of the free electron we adopt a power-law with power index $h=2$ \citep[e.g.][]{2009GrCo...15...20B,er&rogers19}. For the magnetic field, we used the observational result from our Milky Way, which is about $\sim 10\mu$ G on average \citep{2013pss5.book..641B,2017ARA&A..55..111H}. The Larmor frequency is on the level of $\sim$kHz with such a magnetic field. The profile of the magnetic field is a power-law with a different index. We adopt two models in this section ($h_b=1,3$). Such a choice is suggested by observation of the central Black Hole in M87 \citep{2014ApJ...786....5K,2021ApJ...910L..13E}.
We define a scale radius $R_0=10$ kpc, then we can write down the profiles of electron density and magnetic field
\be
n_e(r) = n_0 (R_0/r)^2, \quad B(r) = B_0 (R_0/r)^{h_b}, \quad h_b=1,3
\ee
where we have $n_0=0.01$ cm$^{-3}$ and $B_0=10^{-5}$ Gauss.
We follow the definition of Dispersion Measure (DM) for the approximation of projected electron density 
\be
DM (\theta)\approx N_e(\theta)= \int n_e(r)\, d z.
\ee
and Rotation Measure (RM) for the magnetic field
\be
RM = \frac{2 r_e^2 c \epsilon_0}{e } \int n_e(r) B_\parallel\, d z_l.
\ee
In our choice, we have $DM=100$ pc\,cm$^{-3}$ and $RM=810$ rad\,$m^{-2}$ at $R_0=10$ kpc.
The potential (dispersion) delay can be written by
\be
t_{pl}=\frac{\lambda^2 r_e}{2 \pi c}DM \pm \frac{\lambda^3}{2\pi c} RM.
\ee
Then the plasma lensing potential can be calculated from Eq.\,\ref{eq:pot2dt}, as well as the deflection angle 
\be
\alpha_{pl} =  -\frac{\theta_0^3}{\theta^2} \pm \frac{\theta_B^{h+1}}{\theta^h},
\ee
where $h=2+h_b$, and the characteristic radius is \citep{2010MNRAS.404.1790B,er&rogers19}
\be
\theta_0^3 = \frac{\lambda^2 r_e n_0}{\sqrt{\pi}} \frac{R_0^2}{D_tD_d} \frac{\Gamma(3/2)}{\Gamma(1)},
\ee
and 
\be
\theta_B^{h+1} = \frac{\lambda^3 r_e^2 c \epsilon_0 n_0 B_0}{\sqrt{\pi}e D_t } \, \frac{R_0^h}{D_d^{h-1}} \,  \frac{\Gamma(\frac{h+1}{2})}{\Gamma(h/2)},
\ee
where $\Gamma(x)$ is the Gamma function.

We calculate the deflection angles caused by gravity $\alpha_{gl}$, plasma $\alpha_{pl}$ and magnetic field $\alpha_B$ separately. In a galaxy-scale strong lensing, $\alpha_{gl}$ is on the order of $\sim1$ arcsec. Given the electron density and magnetic field we adopted, $\alpha_{pl}$ is $\sim10^{-4}$ arcsec, and $\alpha_B$ is $\sim10^{-14}$ arcsec for the light at 1 GHz. Thus, the angular separation induced by the magnetic field is too small to be spatially resolved, rendering two images indistinguishable.

\begin{figure}
\includegraphics[width=9cm]{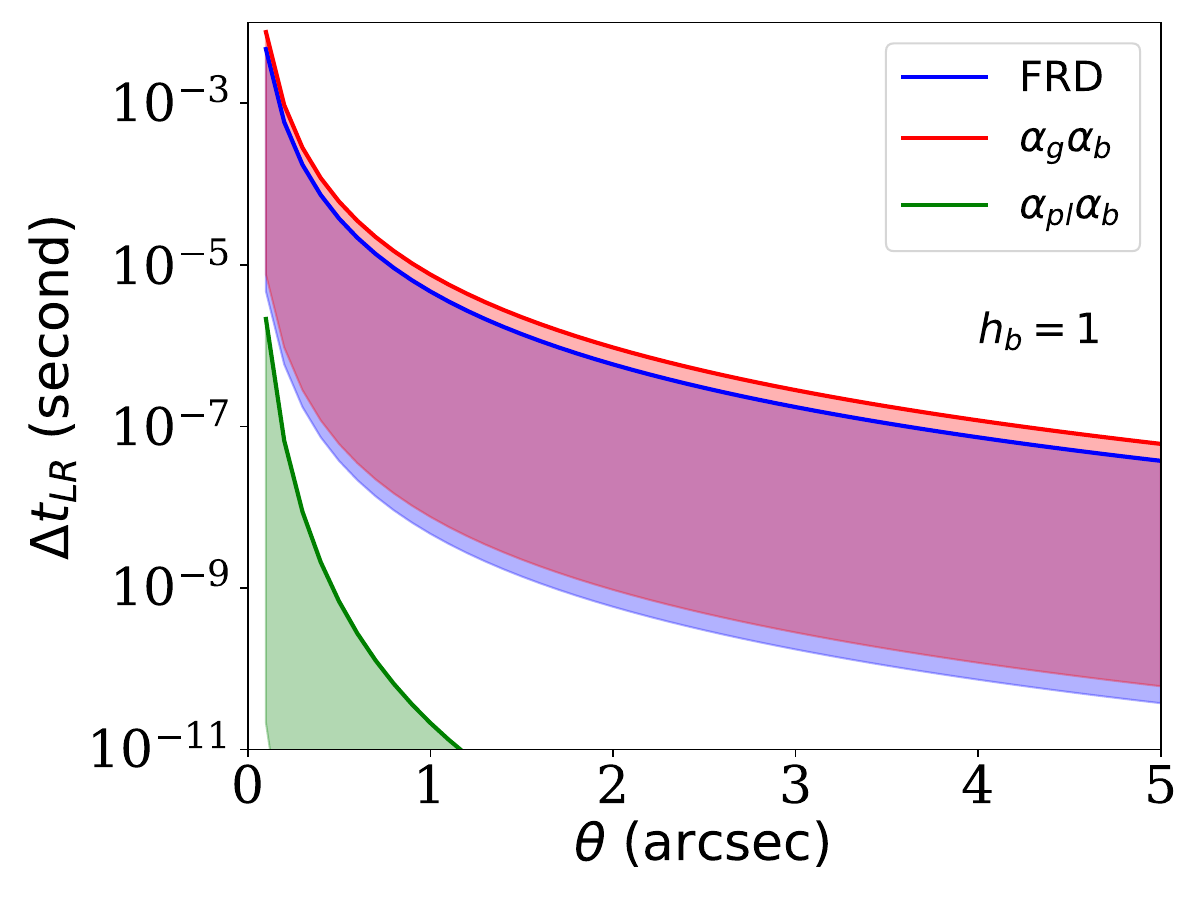}\\
\includegraphics[width=9cm]{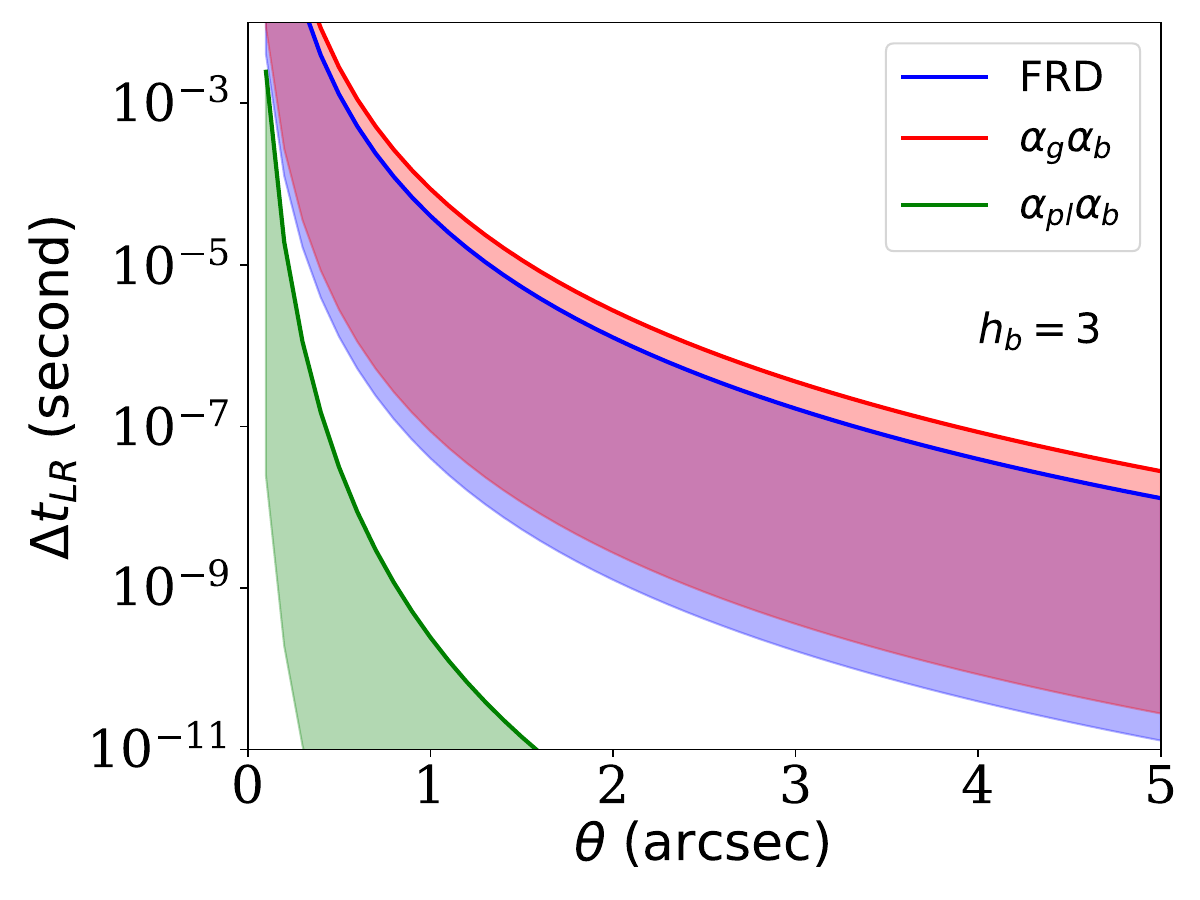}
\caption{The arrival time difference between left- and right-mode polarizations. The shadow covers the frequency $0.5-5$ GHz. The blue shadow presents that due to different velocities of the two modes. The red (green) shadow presents that due to different paths of the propagation caused by gravitational deflection (plasma deflection). The top (bottom) panel is for the magnetic profile with index $h_b=1$ ($h_b=3$).}
\label{fig:dt_sis}
\end{figure}

As mentioned above, the arrival time difference between the two modes is amplified by cosmic distance. In Fig.\,\ref{fig:dt_sis}, we show $\Delta t_{LR}$, the time delay difference as a function of the lensed image position $\theta$. For both magnetic profiles, the delay difference caused by the gravitational deflection (red) is comparable to and greater than those induced by other effects. It exhibits a similar dependence on $\theta$ as the delay associated with Faraday rotation (blue). The delay caused by the plasma deflection, however, is relatively small, and decreases rapidly with increasing $\theta$. 
In the second profile ($h_b=3$), $\Delta t_{LR}$ decreases with $\theta$ rapidly. While at small $\theta$, the delay is large. 
Moreover, for the image formed at small $\theta$, i.e. near the centre of the lens, the arrival time difference is large enough compared to the width of the FRB ($\sim 1$ millisecond). It is possible to widen the shape of the pulse of the signal, or even split the pulse. In gravitational lensing, the images formed within the Einstein radius usually are demagnified. However, a high plasma density in the central region of the lens can mitigate the demagnification, potentially generate detectable images \citep{2022MNRAS.516.2218E}. 

\begin{figure}
\includegraphics[width=9cm]{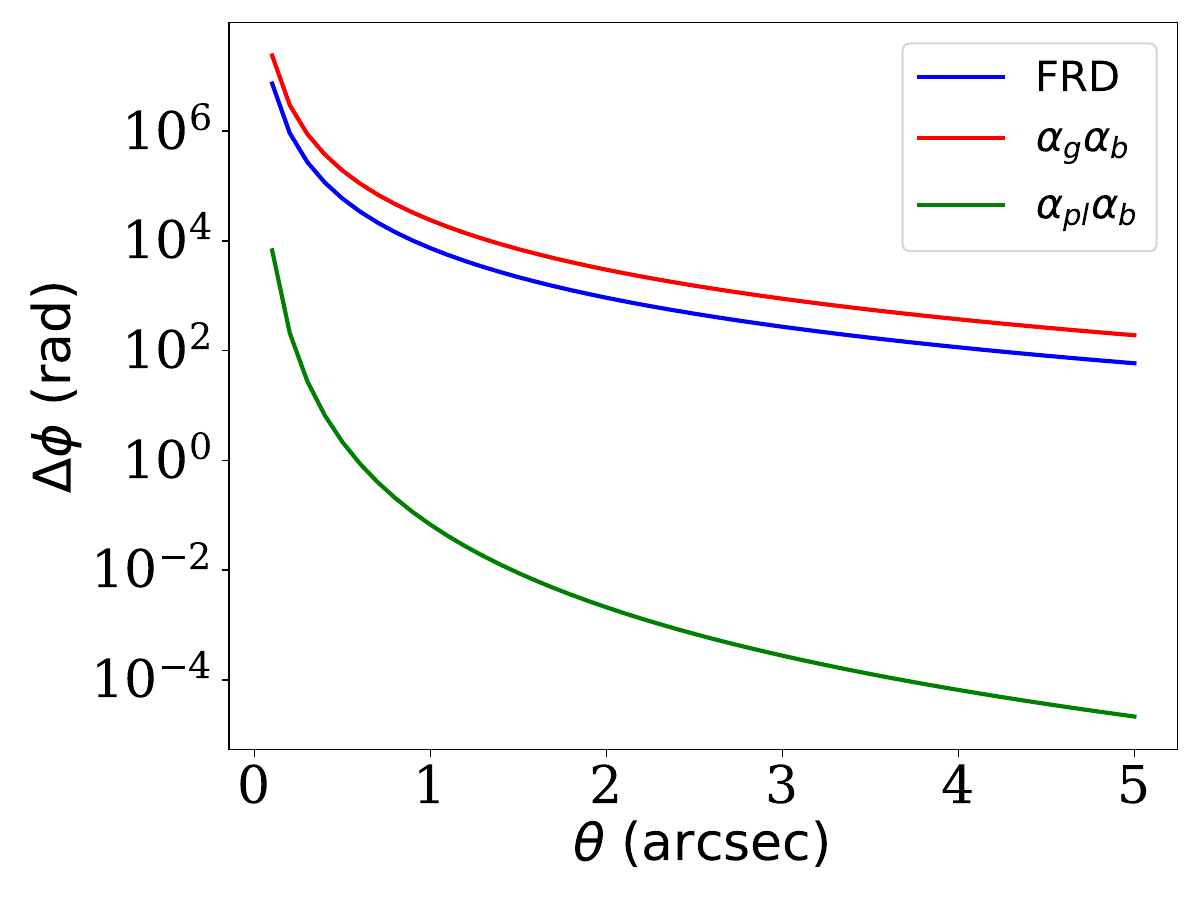}
\includegraphics[width=9cm]{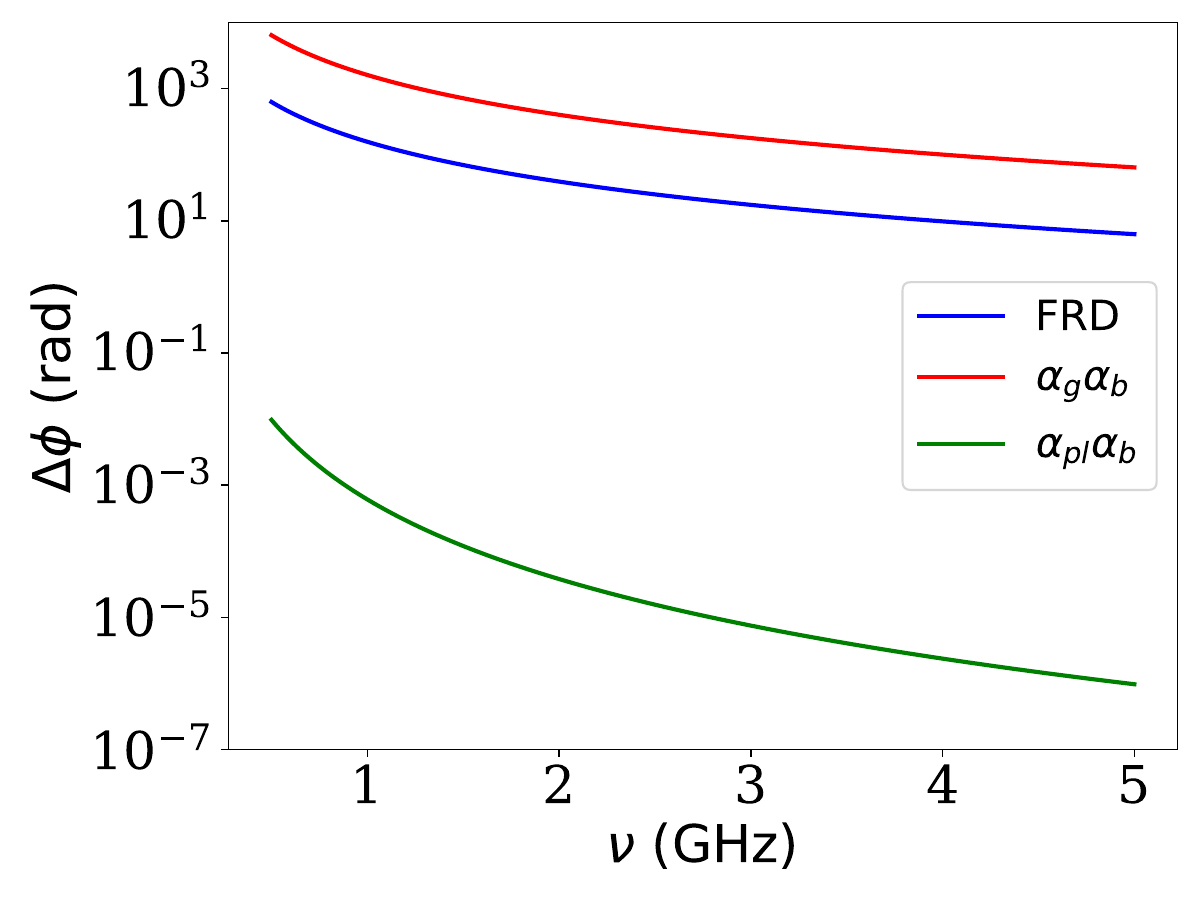}
\caption{The rotation of the linear polarization with image position (top, reference frequency $\omega=0.5$ GHz) and frequency (bottom, image position $\theta=2$ arcsec).}
\label{fig:rotation_sis}
\end{figure}

\begin{figure}
\includegraphics[width=9cm]{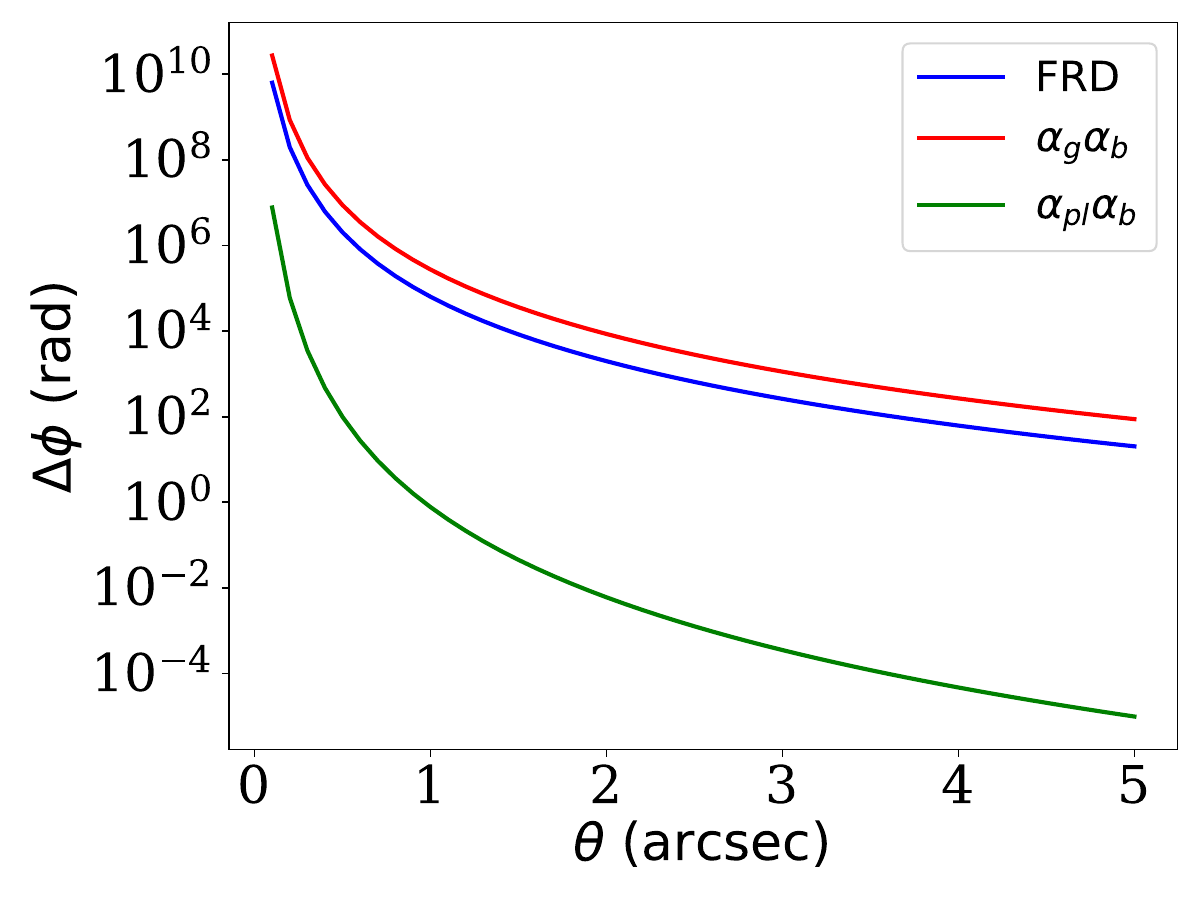}
\includegraphics[width=9cm]{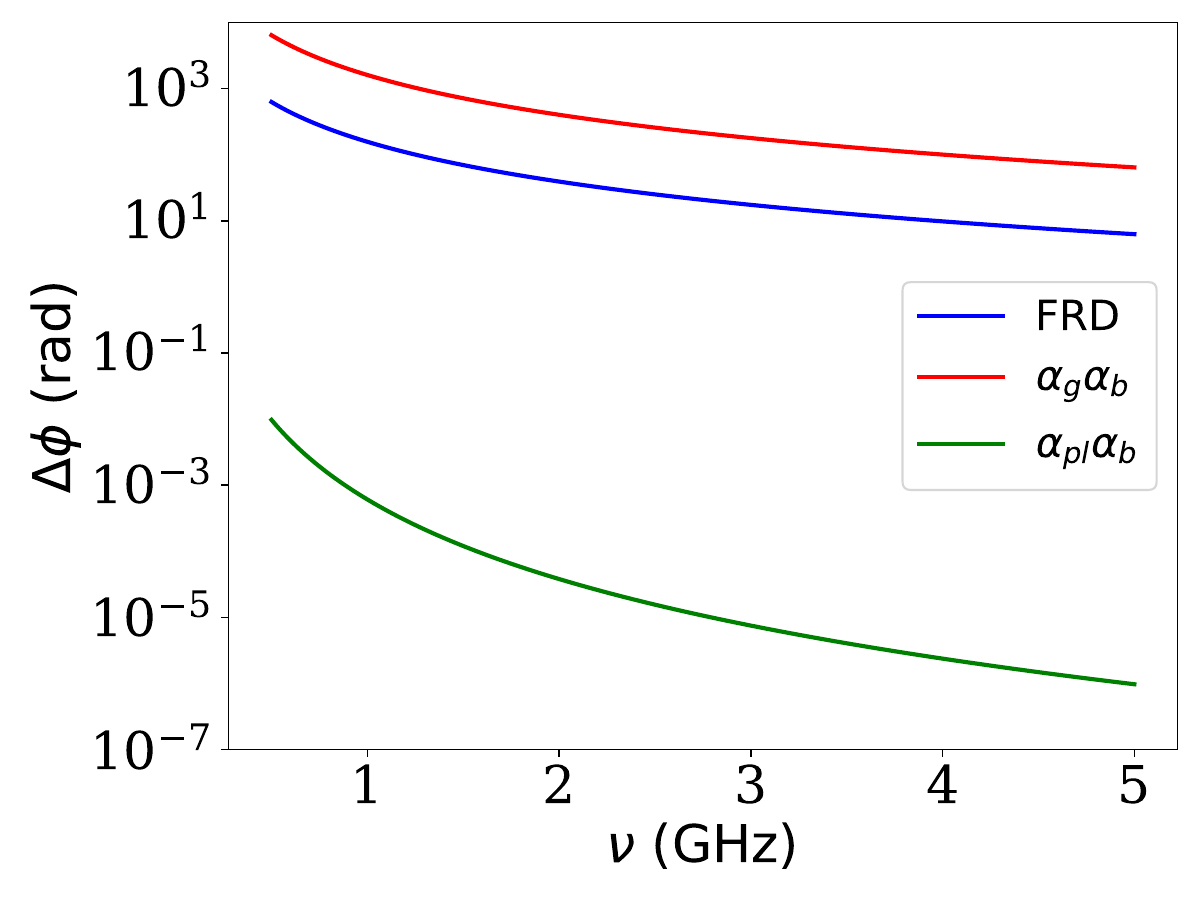}
\caption{Same as Fig.\,\ref{fig:rotation_sis} but for $h_b=3$.}
\label{fig:rotation_sis_hb3}
\end{figure}

In Figs.\,\ref{fig:rotation_sis} \& \ref{fig:rotation_sis_hb3}, the rotations of the linear polarization in a lensed image are shown. In the top panel, it is evident that the rotation is primarily driven by gravitational deflection. The bottom panel illustrates that the rotation due to gravitational deflection exhibits the same frequency dependence as Faraday rotation. Unlike the rotation induced by plasma deflection, which scales as $\propto\nu^{-4}$, the gravitational rotation cannot be distinguished from the Faraday rotation.


\section{Strong magnetic field}
In this section, we consider a case with a relatively strong magnetic field and apply the dispersion relation Eq.\,\ref{eq:dispersion-app}. The Larmor frequency depends on the relative orientation between the light and the magnetic field, which influences the deflection of the light. However, we continue to adopt the thin-lens approximation for simplicity.

We take the point mass model as our example, i.e. a Black Hole, since it has been suggested that in order to launch a jet, it requires strong magnetic field ($\sim 10^{3-4}$ G) near the BH \citep{1977MNRAS.179..433B,2016A&A...593A..47B,2015Sci...350.1242J}. We use a nearby supermassive BH, M87, as our lens, of which the mass is $6\times 10^9$M$\odot$ \citep{2009ApJ...700.1690G}, and the
distance is 16.7 Mpc \citep{2005ApJ...634.1002J}.
The magnetic field model is taken to be a radial power law ($1/r$) again \citep{1974PhRvD..10.3166P}. 
To better demonstrate the effect of the magnetic field, we take a high electron density, $n_e=100$cm$^{-3}$ and a strong magnetic field $B=10$G at $R_0=100r_g$, which agrees with the observation \citep{2014ApJ...786....5K}. The Larmor frequency approaches $\sim0.1$GHz or even higher in the centre of the lens.

\begin{figure}
\centering
\includegraphics[width=10cm]{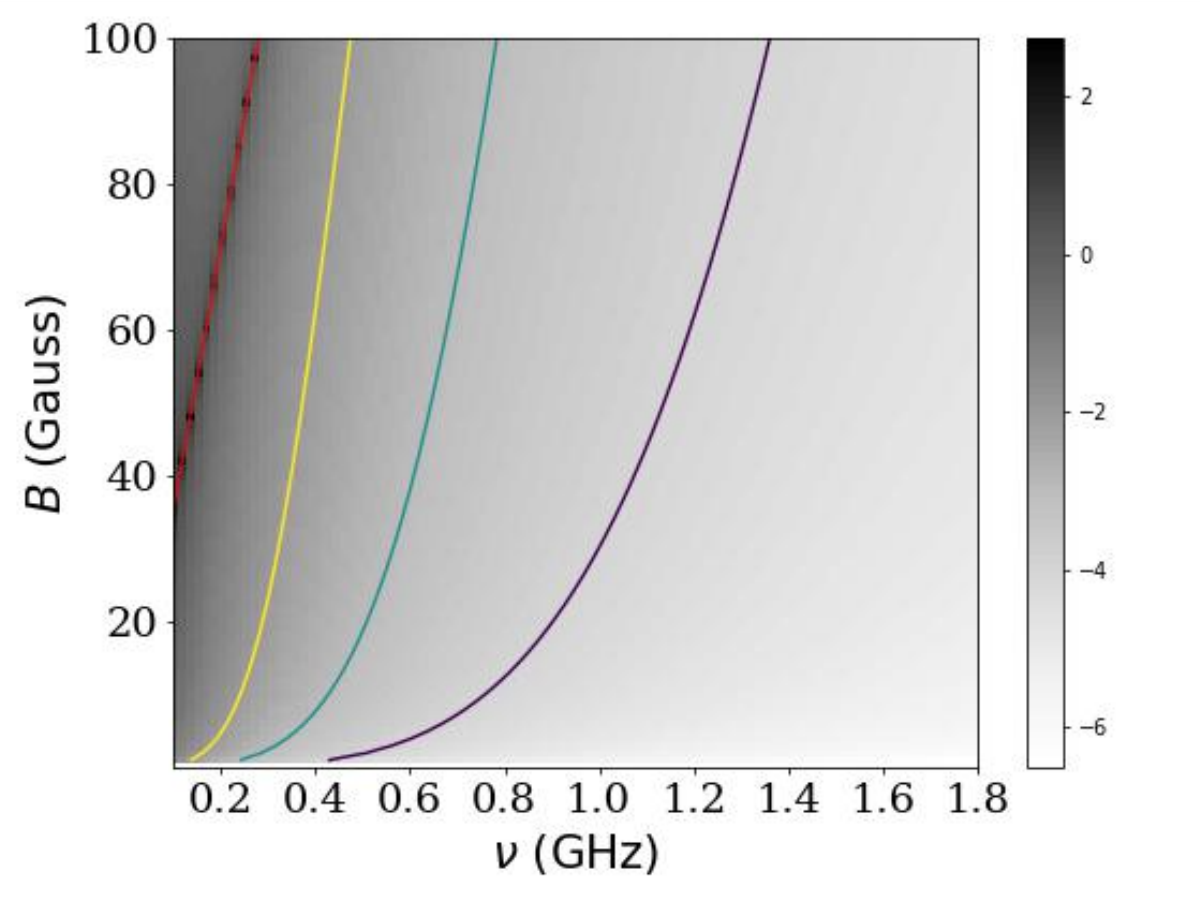}
\caption{The maximum difference of deflection angle between two polarization modes. The point mass model in Section \ref{sec:pointlens} is adopted. The red line shows the condition of image flip. The yellow, green and purple curve marks $\Delta \alpha=10^{-2},10^{-3},10^{-4}$ arcsec respectively. }
\label{fig:image-split}
\end{figure}

The lensing properties of a point mass model with plasma have been studied in \citet{2020MNRAS.491.5636T}, which discusses the hill and hole structures of the magnification curve. Here we compare the difference of the two modes of light. The deflection angle by gravity of a point lens is
\be
\alpha_{gl} = \frac{\theta_E^2}{\theta}, \quad
{\rm with} \quad 
\theta^2_E = \frac{4GM}{c^2}\frac{D_{ds}}{D_dD_s},
\ee
and $M$ is the total mass of the lens. Here, the mass of the free electron is not included, as they will not cause a significant effect. To further simplify the mathematics, we use the source distance with $D_s=2D_d$. The Einstein radius of the point gravitational lens is $\theta_E\approx1.2$ milli-arcsec.

\begin{figure}
\includegraphics[width=9cm]{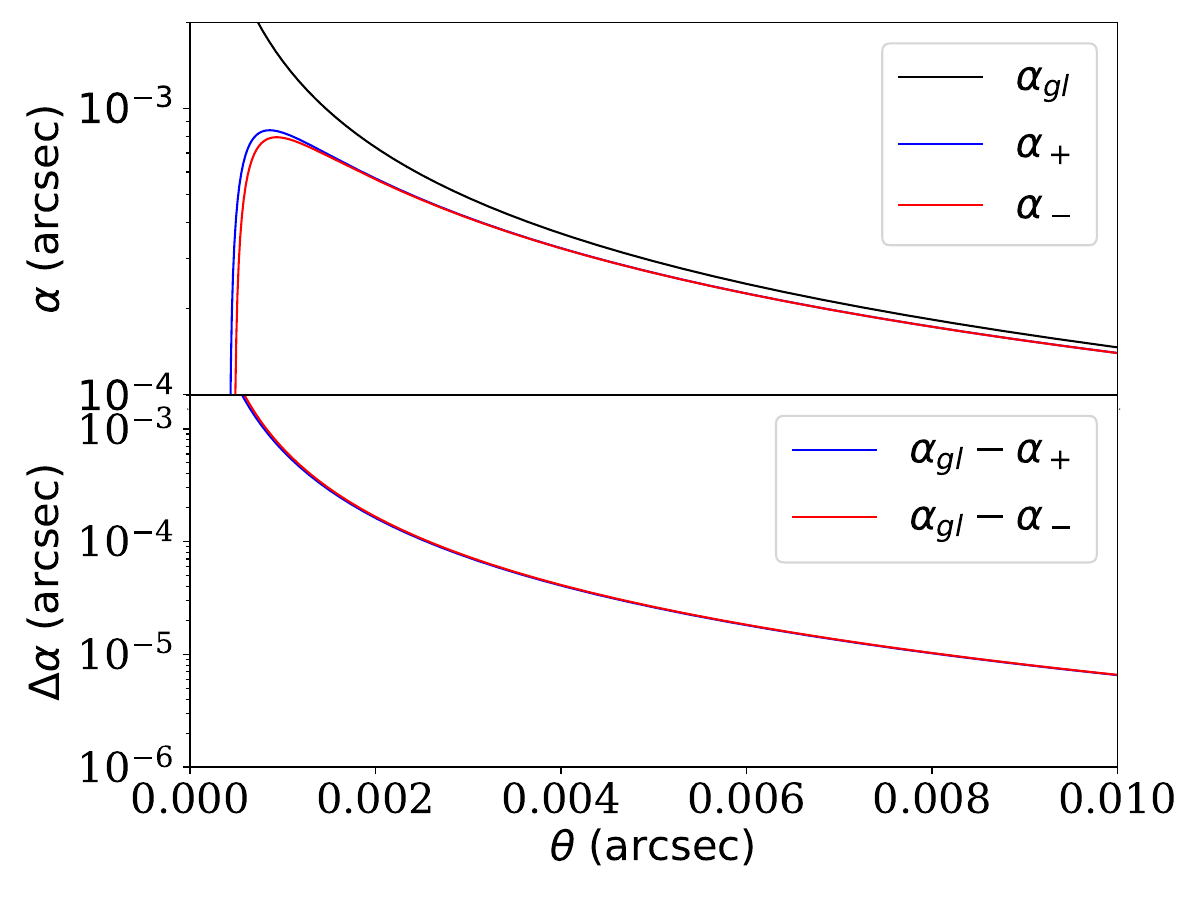}\\
\includegraphics[width=9cm]{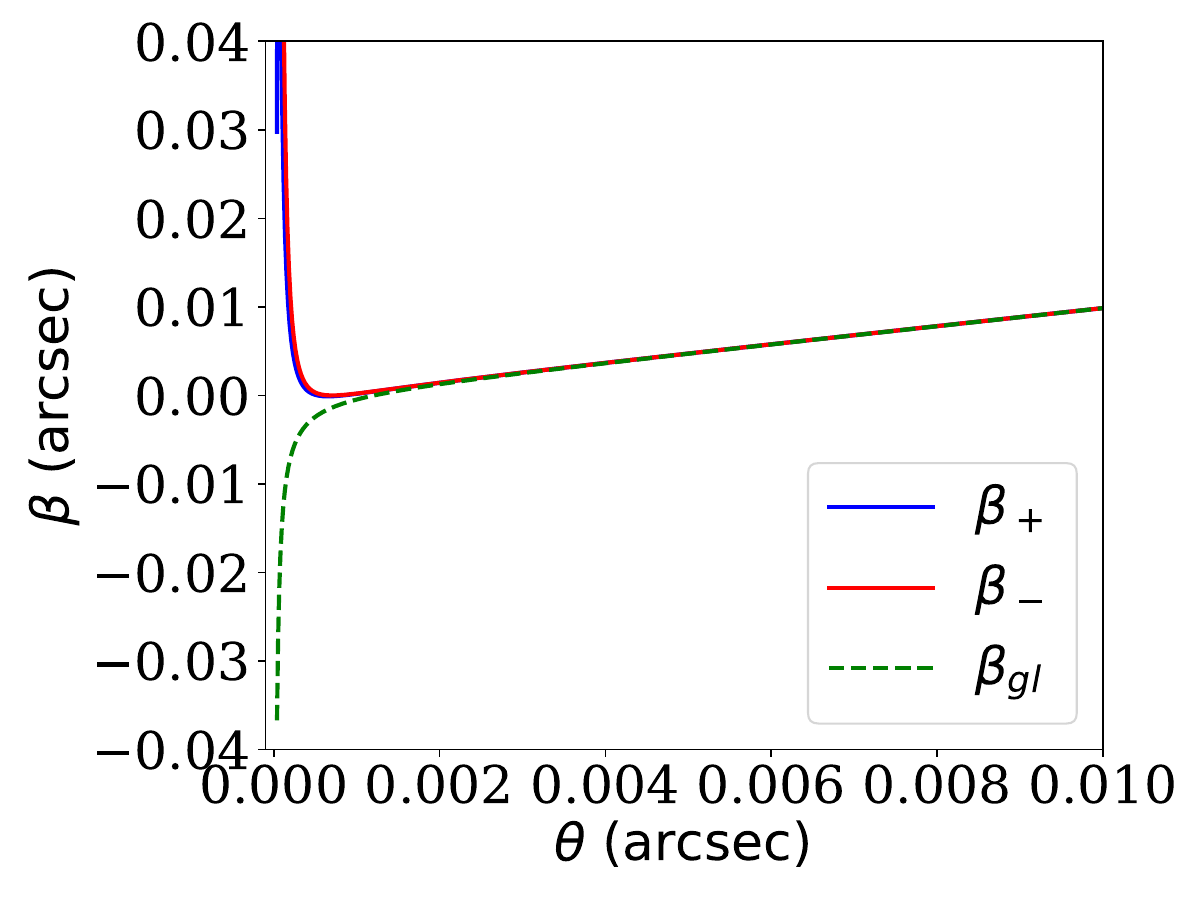}
\caption{$Top-$the deflection angle of a point lens with strong magnetic plasma. The red (blue) curve presents the ordinary (extraordinary) mode of the wave. 
$Bottom-$the relation between source position $\beta$ and image position $\theta$.}
\label{fig:alphapm}
\end{figure}

We calculate the total deflection angles for both polarization modes of light (Fig.\,\ref{fig:alphapm}). Both deflection angles are smaller than those in a vacuum. The difference between the two modes is small, but becomes slightly larger at small $\theta$. While the gravitational deflection in vacuum diverges at the centre of the lens, the deflection angles in the presence of plasma remains finite. In the bottom panel, $\beta_{gl}$ increases monotonically with $\theta$. The combined effects of plasma magnetic deflections alter the behaviours of lensed images near the centre, resulting a pronounced turnover in $\beta_{\pm}$. This turnover is a direct consequence of the deflection induced by the magnetized plasma. 
We compute the image splitting between the two polarization modes, i.e. the difference in the deflection angle $\Delta\alpha$ at $\theta=1$ milli-arcsec. In Fig.\,\ref{fig:image-split}, the gray shading represents $\Delta \alpha$ on a logarithmic scale. The red line indicates the condition where the images will flip between the two polarization modes. Using $0.1$ milli-arcsec as the limit for spatial resolution (purple curve), it is clear that the possibility of detecting two images with different polarizations exists only near the centre of the BH. For this work, we will perform calculations at an observational frequency of $1$ GHz and will exclude image splitting unless otherwise specified.

\begin{figure}
\centering
\includegraphics[width=9cm]{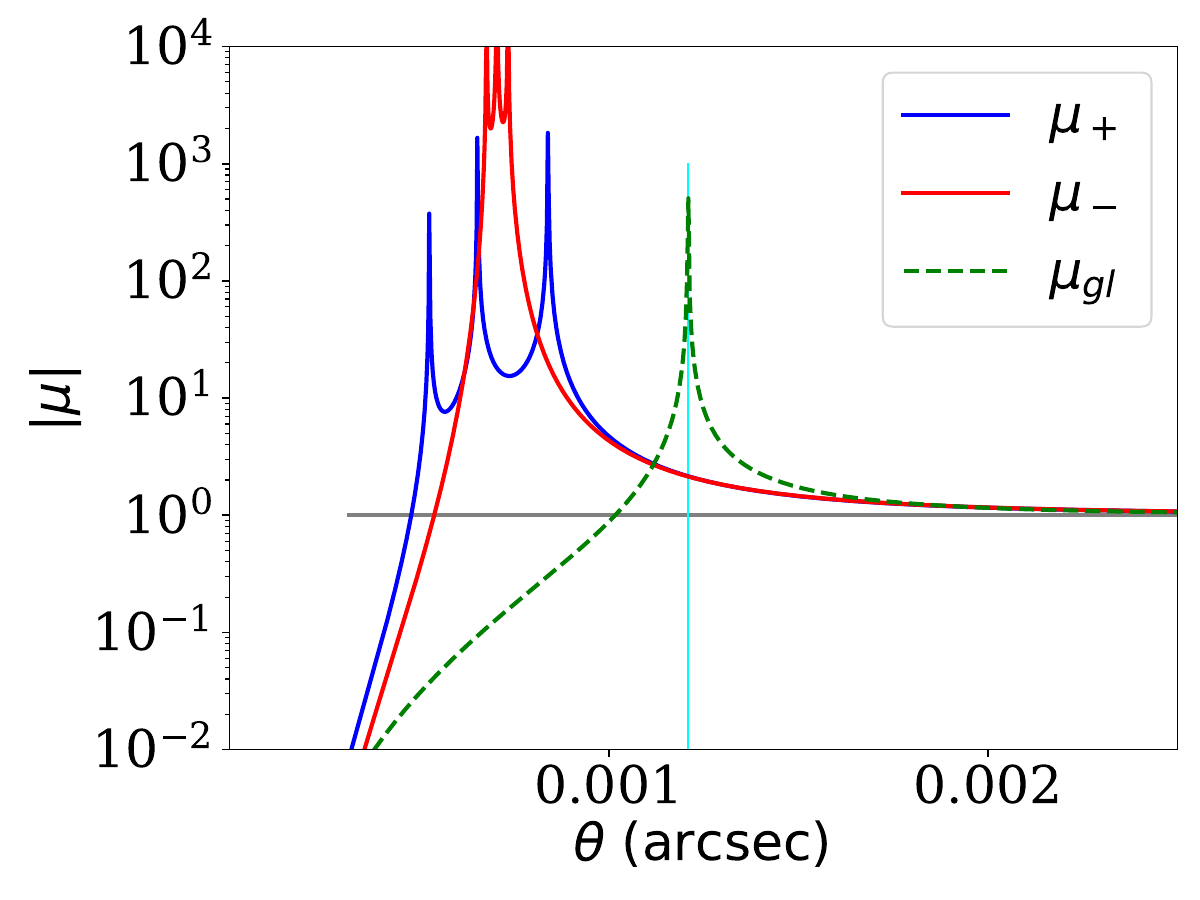}
\caption{The magnification curves of the point lens with magnetic plasma. The green dashed curve presents the magnification in vacuum. The red (blue) one shows that of the ordinary (extraordinary) mode. The cyan vertical line marks the position of the Einstein radius in vacuum.}
\label{fig:mu_pt}
\end{figure}

\begin{figure}
\centering
\includegraphics[width=9cm]{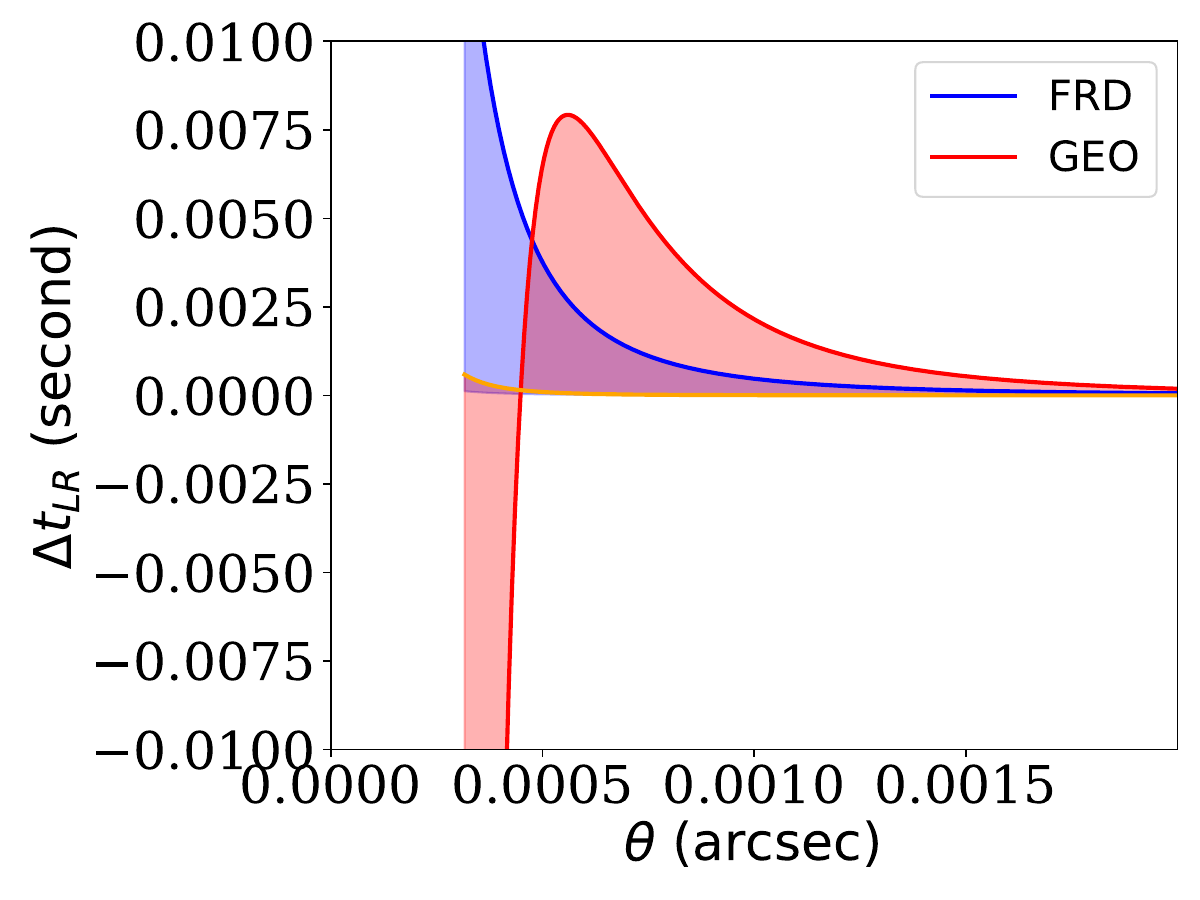}
\caption{Similar to Fig.\,\ref{fig:dt_sis}: the red colour presents the arrival time difference between the two modes of the wave caused by the different paths of the light propagation ($t_{\rm extraordinary}-t_{\rm ordinary}$). The shadows cover a frequency range from 1 to 5 GHz.}
\label{fig:dt_pt}
\end{figure}

\begin{figure}
\includegraphics[width=9cm]{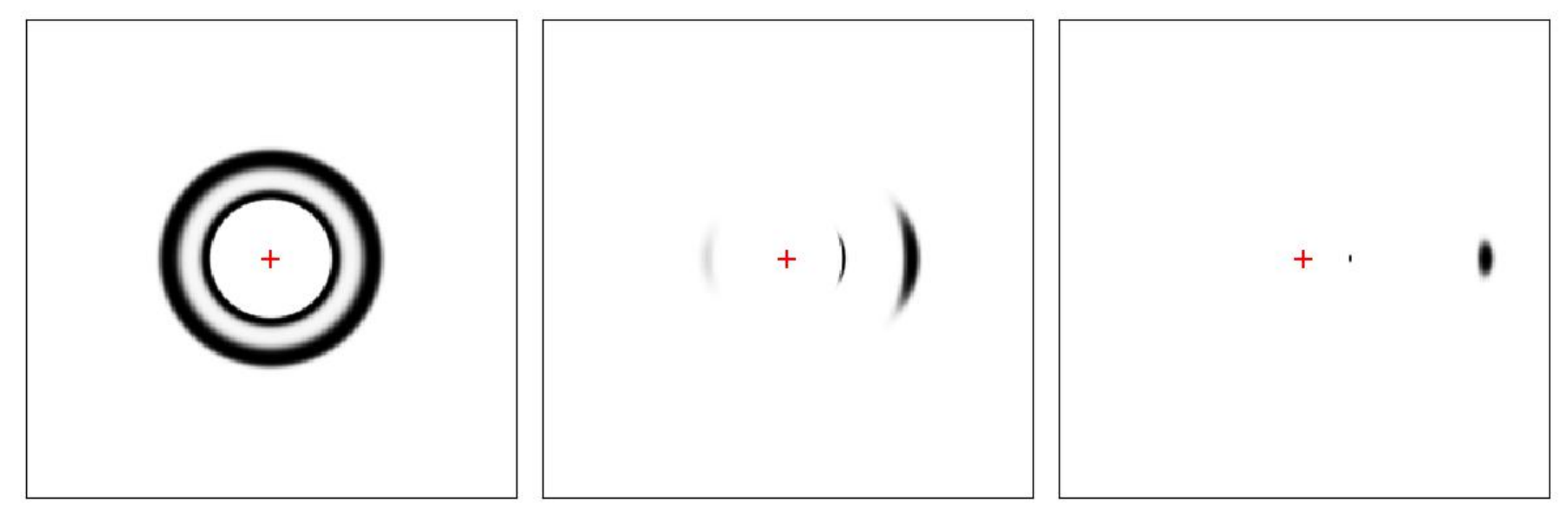}\\
\includegraphics[width=9cm]{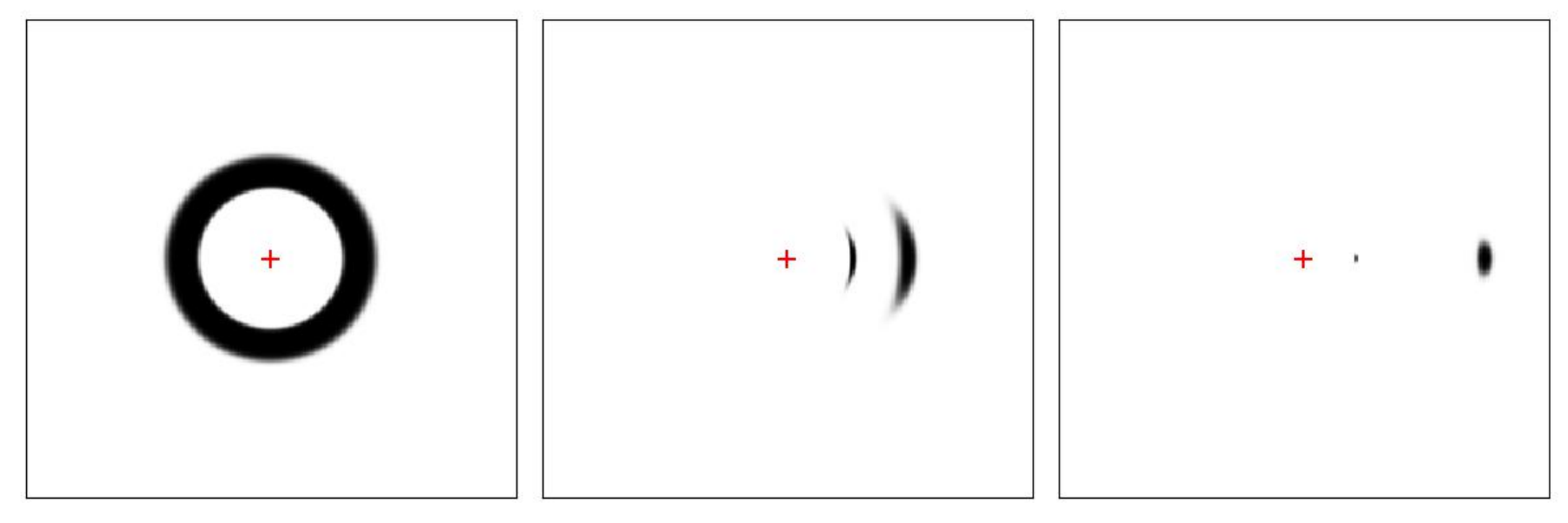}
\caption{Demonstration of simulated lensed images. The red plus in the centre marks the position of the lens. Top: only gravitational deflection; middle: the deflection with magnetic plasma for extraordinary mode; bottom: the deflection with magnetic plasma for ordinary mode. The Einstein radius of the point-lens is $\sim1.2$ milli-arcsec. The relative position of the background source with respect to the lens is $0,0.5,1,1.5,2$ milli-arcsec from left to right respectively.}
\label{fig:images}
\end{figure}

In Fig.\,\ref{fig:mu_pt}, we compare the magnification curves for a lens in a magnetic plasma with those in a vacuum, focus on the central region of the lens. The critical radius (where $\mu\to\infty$) is reduced for both polarization modes. Moreover, both modes exhibit extra hill and hole structure in the inner region, resembling the result in plasma lensing \citep[e.g.][]{2020MNRAS.491.5636T}. These structures imply the potential formation of multiple ``Einstein ring'', although detecting them would require exceptionally high resolution.
We show the arrival time difference between the two polarization modes in Fig.\,\ref{fig:dt_pt}. The blue shading represents the delay caused by the differing velocities of the two modes, while the red shading indicates the delay due to the difference in their trajectories. An interesting point is that for images formed close to the centre of the lens, the time difference reverses. This reversal can be interpreted as an inversion in dispersion relation, as we discussed in Eq.\,\ref{eq:dispersion-compact}. In this scenario, the trajectory difference becomes the dominant factor influencing the delay time. Therefore, the polarization information is determined by both the deflection of the signal and the magnetic field. 
However, the time delays presented here are calculated based on the positions of lensed images. In reality, the lensed images corresponding to the two modes originate from slightly different source positions. This discrepancy becomes significant in the presence of a strong magnetic field. Although we do not consider the image splitting in this work, we provide an example of lensed images for two polarization modes in Fig.\,\ref{fig:images}. For the ordinary mode ($+$mode), a ``double Einstein ring'' can form. In the top middle panel, we can also see that an extra faint image can form on the opposite side of the lens.

For the magnetic profile with power index $h_b=3$, similar lensing behaviour can be found but rapid decreasing with $\theta$. Complex lensing effects will appear near the centre of the lens. Such a strong magnetic field usually can be only found near a neutron star or black hole, as shown by the Event Horizon Telescope near M87 \citep{2021ApJ...910L..13E}. The observation of M87 may not be dramatically affected by lensing effects since the lensing distance $D_t$ is small. The magnetic field and its gradient is strong and may affects the Faraday rotation relation.


\section{Summary and discussions}

In this work, we study the gravitational lensing embedded in a magnetized plasma. The magnetized medium introduces additional deflections and exhibits birefringence. In a uniform magnetic field, the two polarization modes propagate along an identical path but with different arrival times or phases, which causes the Faraday rotation of the linear polarization. However, in the presence of magnetic field gradients, the trajectories of the two polarization modes diverge. The distinction of the light path is further enhanced by the gravitational field, resulting in significant time delays (which we called geometric delay) and additional effects, such as rotation of the linear polarization.
First, we compare the two modes in a galaxy-scale lens with a weak magnetic field. Geometric delay and geometric rotation show the same dependence on observational frequency as Faraday rotation, and their magnitudes are greater. This lensing induced geometric rotation occurs generically in the presence of magnetic field gradient and plasma density gradient. For an order of magnitude estimate, only strongly lensed sources, i.e. deflection angle $\alpha\sim1$ arcsec, at cosmic distance, $D_t\sim100$ Mpc, the geometric rotation is comparable or even greater than the Faraday rotation.
In the second example, we consider a point lens with a strong magnetic field. The lensed images in plasma differ from those in vacuum in several ways. The deflection angle is smaller in plasma medium than that in vacuum. The hill and hole structures can appear in the magnification curve, varying between the two polarization modes. The lensed images can split for the two modes, displaying complex behaviours, especially near the centre of the lens, although extremely high resolution is required to resolve this. For short period radio sources, such as FRBs, the distinct time delays between the two modes may alter the pulse structure, potentially leading to multiple pulses with different polarization modes. 

The study presented here is based on several approximations in the calculations, and requires further investigations. First of all, the refraction index employed is a scaler and thus insufficient to describe all propagation modes of the light, especially under the conditions involving large deflection angles. In case of a strong gravitational field, e.g. near a black hole, the small angle approximation breaks down. The lens equation describing the deflection in the strong field incorporating tensor mode dispersion relation is necessary to model these effects. The photon frequency will be changed by the strong gravitational field, and also affect the results in this work. Additionally, while the thin-lens approximation adopted in this study is adequate for our example of a galaxy lens, it requires modification when applied to more complex scenarios. 

The temperature of the plasma in the interstellar medium or intergalactic medium may not be a critical factor to affect the results in this study. The situation will not be the same when we look at the vicinity of BH. The deflection can be complicated because of not only strong gravity, but also the high-temperature gas.
Moreover, since the difference between the two modes is small, other effects could influence the result, such as the motion of the lens and source, the structures along the line-of-sight, the variations/small structures in plasma density etc. Nevertheless, polarization measurements offer valuable additional constraints on both the lens system and the properties of the background source itself. The propagation effects, particularly in regions influenced by strong gravitational field, are non-negligible and must be explicitly accounted for.

\begin{acknowledgements}
I thank the referee for constructive and valuable comments. I also like to thank Oleg Yu Tsupko for valuable suggestions and comments on the work.
\end{acknowledgements}

\bibliographystyle{aa} 
\bibliography{bfield} 
\end{document}